\documentstyle[12pt]{article}
\input epsf.sty
\catcode`@=11
\def\chkspace{%
  \relax   
  \begingroup\ifhmode\aftergroup\dochksp@ce\fi\endgroup}
\def\dochksp@ce{%
  \unskip              
  \futurelet\chkspct@k\d@chkspc  
}
\def\d@chkspc{%
  \let\nxtsp@ce=\relax
  \ifx\chkspct@k.\else     
    \ifx\chkspct@k,\else
      \ifx\chkspct@k;\else
        \ifx\chkspct@k!\else
          \ifx\chkspct@k?\else
            \ifx\chkspct@k:\else
              \ifx\chkspct@k)\else
              \ifx\chkspct@k(\else
                \ifx\chkspct@k]\else
                  \ifx\chkspct@k-\else
                    \ifx\chkspct@k\egroup\else  
                      \let\nxtsp@ce=\put@space  
                    \fi
                  \fi
                \fi
              \fi
              \fi
            \fi
          \fi
        \fi
      \fi
    \fi
  \fi
  \nxtsp@ce
}
\def\put@space{$\;$}
\catcode`@=12

\def\ra{\relax\ifmmode \rightarrow\else{{$\rightarrow$}}\fi\chkspace}
\def\Ra{\relax\ifmmode \Rightarrow\else{{$\Rightarrow$}}\fi\chkspace}

\def\eg{{\it eg.}\chkspace}

\def\ep{{e$^+$e$^-$}\chkspace}

\def\qu{\relax\ifmmode \quad\else{{$\quad$}}\fi\chkspace}

\def\gluino{\relax\ifmmode \tilde{g} \else $\tilde{g}$ \fi\chkspace}

\def\qq{\relax\ifmmode q\overline{q}
\else $q\overline{q}$ \fi\chkspace}
\def\ff{\relax\ifmmode f\overline{f}
\else $f\overline{f}$ \fi\chkspace}

\def\bb{\relax\ifmmode b\bar{b}
       \else $b\bar{b}$ \fi\chkspace}
\def\ccrm{\relax\ifmmode {\rm c}\bar{\rm c}
       \else ${\rm c}\bar{\rm c}$ \fi\chkspace}
\def\cc{$c\bar{c}$ \chkspace}
\def\tt{\relax\ifmmode {\rm t}\bar{\rm t}
       \else ${\rm t}\bar{\rm t}$ \fi\chkspace}
\def\ss{\relax\ifmmode {\rm s}\bar{\rm s}
       \else ${\rm s}\bar{\rm s}$ \fi\chkspace}
\def\uu{\relax\ifmmode {\rm u}\bar{\rm u}
       \else ${\rm u}\bar{\rm u}$ \fi\chkspace}
\def\dd{\relax\ifmmode {\rm d}\bar{\rm d}
       \else ${\rm d}\bar{\rm d}$ \fi\chkspace}

\def\qqg{\relax\ifmmode q\overline{q}g
\else $q\overline{q}g$ \fi\chkspace}
\def\bbg{\relax\ifmmode b\overline{b}g
\else $b\overline{b}g$ \fi\chkspace}
\def\ccg{\relax\ifmmode c\overline{c}g
\else $c\overline{c}g$ \fi\chkspace}
\def\ttg{\relax\ifmmode t\overline{t}g
\else $t\overline{t}g$ \fi\chkspace}

\def\afb{\relax\ifmmode A_{FB} \else
{{$A_{FB}$}}\fi\chkspace}
\def\afbb{\relax\ifmmode A_{FB}^b \else
{{$A_{FB}^b$}}\fi\chkspace}
\def\pafb{\relax\ifmmode \tilde{A}_{FB} \else
{{$\tilde{A}_{FB}$}}\fi\chkspace}
\def\pafbb{\relax\ifmmode \tilde{A}_{FB}^b \else
{{$\tilde{A}_{FB}^b$}}\fi\chkspace}

\def\pafbzo{\relax\ifmmode \tilde{A}_{FB}|_{O(0)} \else
{{$\tilde{A}_{FB}|_{O(0)}$}}\fi\chkspace}
\def\pafbfo{\relax\ifmmode \tilde{A}_{FB}|_{\oalp} \else
{{$\tilde{A}_{FB}|_{\oalp}$}}\fi\chkspace}
\def\pafbso{\relax\ifmmode \tilde{A}_{FB}|_{\oalpsq} \else
{{$\tilde{A}_{FB}|_{\oalpsq}$}}\fi\chkspace}
\def\pafbto{\relax\ifmmode \tilde{A}_{FB}|_{\oalpc} \else
{{$\tilde{A}_{FB}|_{\oalpc}$}}\fi\chkspace}

\def\pafbbzo{\relax\ifmmode \tilde{A}_{FB}^b|_{O(0)} \else
{{$\tilde{A}_{FB}^b|_{O(0)}$}}\fi\chkspace}
\def\pafbbfo{\relax\ifmmode \tilde{A}_{FB}^b|_{\oalp} \else
{{$\tilde{A}_{FB}^b|_{\oalp}$}}\fi\chkspace}
\def\pafbbso{\relax\ifmmode \tilde{A}_{FB}^b|_{\oalpsq} \else
{{$\tilde{A}_{FB}^b|_{\oalpsq}$}}\fi\chkspace}
\def\pafbbto{\relax\ifmmode \tilde{A}_{FB}^b|_{\oalpc} \else
{{$\tilde{A}_{FB}^b|_{\oalpc}$}}\fi\chkspace}

\def\afbo0{\tilde{A}_{FB}|_{O(0)}}
\def\afbo1{\tilde{A}_{FB}|_{\oalp}}
\def\afbo2{\tilde{A}_{FB}|_{\oalpsq}}
\def\afbo3{\tilde{A}_{FB}|_{\oalpc}}

\def\lam{\relax\ifmmode \Lambda_{\overline{MS}}
       \else {{$\Lambda_{\overline{MS}}$}}\fi\chkspace}
\def\lamuds{\relax\ifmmode \Lambda^{(3)}_{\overline{MS}}
       \else {{$\Lambda^{(3)}_{\overline{MS}}$}}\fi\chkspace}
\def\lamudsc{\relax\ifmmode \Lambda^{(4)}_{\overline{MS}}
       \else $\Lambda^{(4)}_{\overline{MS}}$\fi\chkspace}
\def\lamudscb{\relax\ifmmode \Lambda^{(5)}_{\overline{MS}}
       \else $\Lambda^{(5)}_{\overline{MS}}$\fi\chkspace}

\def\alp{\relax\ifmmode \alpha_s\else $\alpha_s$\fi\chkspace}
\def\alpbar{\relax\ifmmode \bar{\alpha_s}
       \else $\bar{\alpha_s}$\fi\chkspace}
\def\alpmz{\relax\ifmmode \alpha_s(M_Z)\else $\alpha_s(M_Z)$\fi\chkspace}
\def\alpmzsq{\relax\ifmmode \alpha_s(M_Z^2)
       \else $\alpha_s(M_Z^2)$\fi\chkspace}

\def\oalp{\relax\ifmmode O(\alpha_s)\else{{O($\alpha_s$)}}\fi\chkspace}
\def\oalpsq{\relax\ifmmode O(\alpha_s^2)
           \else{{O($\alpha_s^2$)}}\fi\chkspace}
\def\oalpc{\relax\ifmmode O(\alpha_s^3)
           \else{{O($\alpha_s^3$)}}\fi\chkspace}
\def\oalpf{\relax\ifmmode O(\alpha_s^4)
           \else{{O($\alpha_s^4$)}}\fi\chkspace}

\def\rb{\relax\ifmmode R_3^b/R_3^{all}
           \else{{$R_3^b/R_3^{all}$}}\fi\chkspace}
\def\rc{\relax\ifmmode R_3^c/R_3^{all}
           \else{{$R_3^c/R_3^{all}$}}\fi\chkspace}
\def\ruds{\relax\ifmmode R_3^{uds}/R_3^{all}
           \else{{$R_3^{uds}/R_3^{all}$}}\fi\chkspace}
\def\ri{\relax\ifmmode R_3^i/R_3^{all}
           \else{{$R_3^i/R_3^{all}$}}\fi\chkspace}
\def\rj{\relax\ifmmode R_3^j/R_3^{all}
           \else{{$R_3^j/R_3^{all}$}}\fi\chkspace}
\def\alpi{\relax\ifmmode \alpha^i_s/\alpha^{all}_s
           \else{{$\alpha^i_s/\alpha^{all}_s$}}\fi\chkspace}

\def\mbz{\relax\ifmmode m_b(M_Z)
           \else{{$m_b(M_Z)$}}\fi\chkspace}
\def\mbb{\relax\ifmmode m_b(M_b)
           \else{{$m_b(M_b)$}}\fi\chkspace}

\def\z0{\relax\ifmmode Z^0 \else {$Z^0$} \fi\chkspace}
\def\h0{\relax\ifmmode H^0 \else {$H^0$} \fi\chkspace}
\def\Dst{\relax\ifmmode {\rm D}^* \else {D$^*$}\fi\chkspace}
\def\Dpl{\relax\ifmmode {\rm D}^+ \else {D$^+$}\fi\chkspace}
\def\D0{\relax\ifmmode {\rm D}^0 \else {D$^0$}\fi\chkspace}
\def\Kst{\relax\ifmmode {\rm K}^* \else {K$^*$}\fi\chkspace}
\def\K0{\relax\ifmmode {\rm K}^0_s \else {K$^0_s$}\fi\chkspace}
\def\Kpl{\relax\ifmmode {\rm K}^+ \else {K$^+$}\fi\chkspace}
\def\Kstz{\relax\ifmmode {\rm K}^{*0} \else {K$^{*0}$}\fi\chkspace}

\def\beq{\begin{equation}}
\def\eeq{\end{equation}}
\def\bea{\begin{eqnarray}}
\def\eea{\end{eqnarray}}

\begin{document}

{\hfill{OUNP-99-13}}

{\hfill{October 1999}}

\vskip 2truecm

\begin{center}

\Large\bf

A Detector for Precision Study of\\ High Energy \ep Annihilations:\\ 
The ECFA/DESY Design for TESLA

\vskip 1truecm

\large

P N Burrows$^{\dag}$

\vskip .6truecm

Particle \& Nuclear Physics, Keble Rd., Oxford, OX1 3RH, UK\\
E-mail: {p.burrows@physics.ox.ac.uk}\\
             
\vskip 1truecm

{\it Representing the ECFA/DESY Study Groups}

\vskip 2truecm

{\it Talk presented at the International Europhysics Conference on High Energy Physics,
Tampere, Finland, 15-21 July 1999.}

\end{center}

\vfill
{$^{\dag}$ Supported by the UK Particle Physics \& Astronomy Research Council}
\eject

\section{Physics at a High Energy Linear Collider}

A high-energy linear \ep collider designed to operate in the c.m. energy
range around and above 500 GeV is an obvious next step for particle physics
investigations of the origin of mass and the mechanism of electroweak
symmetry-breaking, and searches for new dynamics such as Supersymmetry (SUSY). 
The collider represents a natural facility for  discovery and
precision measurement of new particles~\cite{cdr}, and would complement
the physics potential of the LHC. 
For example, for $M_H\sim 100$GeV, favoured by current data, 
tens of thousands of clean \h0 bosons per year would be delivered at
design luminosity.

The anticipated event sample comprises:

\noindent
$\bullet$
Multijet states containing heavy flavours, \eg

\ep \qu \ra \qu \z0 $H^0$ \qu  \ra \qu \ff  \bb/\cc/$\tau^+\tau^-$

\ep \qu \ra \qu \tt \qu \qu \qu \ra \qu $b W^+$ $\bar{b} W^-$

\ep \qu \ra \qu \tt $H^0$ \qu $\;\;$ \ra \qu $b W^+$ $\bar{b} W^-$ \bb

\ep \qu \ra \qu $H^0$ $A^0$ \qu $\;$  \ra \qu \tt \tt

\ep \qu \ra \qu $\tilde{t}\,\tilde{\bar{t}}$ \qu \qu $\;\;\;$ \ra \qu
$\tilde{\chi^0}c\,\tilde{\chi^0} \bar{c} $

\noindent
Events such as these require a high-resolution tracking system
with excellent secondary decay vertex resolution.
Jet energies will typically be in the range 50 \ra 200 GeV, and
the track momentum distribution will peak around 2 GeV/$c$,
so multiple scattering will be important and necessitate low-mass
tracking detectors.

\noindent
$\bullet$
Events with missing energy, \eg 

\ep \qu \ra \qu $\tilde{l^+}\,\tilde{l^-}$ or $\tilde{q}\,\tilde{\bar{q}}$

\ep \qu \ra \qu $\tilde{\chi^+}\,\tilde{\chi^-}$ or $\tilde{\chi^0}\,\tilde{\chi^0}$ 

\noindent
$\bullet$
and perhaps exotic processes, \eg

\ep \qu \ra \qu $\tilde{\chi^0}\,\tilde{\chi^0}$ \qu \ra \qu 
$\tilde{G} \, \gamma\;\tilde{G} \, \gamma$

\ep \qu \ra \qu $G \; \gamma$ 

\noindent
due to gauge-mediated SUSY breaking and extra compact dimensions, respectively. 
Such signatures demand a hermetic calorimeter with good
energy resolution and high granularity for 
energy-flow measurement. This will allow precise jet-jet invariant mass
determination and reconstruction of new heavy states above a large 
combinatorial background. Exotic signatures comprising photons and 
large missing energy also motivate consideration of a 
continuous event readout mode with a software trigger~\cite{guenter}.

\section{Accelerator and Detector Environment}
\vspace*{-.2cm}
The TESLA collider~\cite{cdr,delahaye}, utilising superconducting RF 
cavities for the
main linac, is being designed by an international consortium based around DESY.
The collider operates in a `one-shot' mode at a frequency of 5 Hz. In each cycle
a train of 2820 $5\times550$nm$^2$ e$^-$ bunches meets a similar e$^+$ bunch-train, 
with a bunch separation of 337 ns. The resulting backgrounds
for the detector require careful planning. 
For example, at $\sqrt{s}$ = 500 GeV one expects, {\it per bunch crossing}:

\noindent
$\bullet$
$\sim$ 120k \ep \Ra a large detector B-field;

\noindent
$\bullet$
$\sim$ 1000 $\gamma$ in tracking volume
$\Rightarrow$ highly granular tracking system,
and possible bunch tagger with time resolution $\leq$ 100ns;

\noindent
$\bullet$
several TeV EM energy in the forward regions: $\theta$ $<$ 100 mrad \Ra
shielding and masking;

\noindent
as well as $\sim$ 10$^9$ neutrons/cm$^2$/year, requiring shielding of the 
inner detector.

\section{Overview of Detector Design}
\vspace*{-.2cm}
A schematic of the current design is shown in Fig.~1.
The general concept is a large detector with a gaseous main tracking
chamber and a hermetic, highly granular calorimeter.
A first iteration was presented in~\cite{cdr}.
The design is evolving, and R\&D is underway in all
areas, but the options and
technology choices are being focussed and refined. A brief summary
of the current thinking is given below.

\subsection{Vertex Detector (VXD)} 
\vspace*{-.2cm}
The requirements are high granularity (for low occupancy),
low mass (for low multiple scattering), good spatial resolution (for
precise vertex-finding), and neutron radiation tolerance (see above).
A multi(4 or 5)-layer `self-tracking'
device would be optimal. CCDs and
LHC-style active pixel sensors (APS) are being considered; all
are radiation hard at the expected level. Large-area CCD arrays have been
`combat tested' at SLD/SLC and offer $<$4$\mu$m space-point resolution
in 20$\times$20$\mu$m$^2$ pixels, with the possibility of devices as
thin as 0.12\% $X_0$/layer. The APS pixels will be larger
($50\times50\mu$m$^2$) and thicker (0.8\% $X_0$/layer), but are more
radiation tolerant. CMOS pixel devices have also been suggested and
may allow CCD-like resolution with higher radiation tolerance.
A 3 or 4 T solenoidal magnetic field would confine most of the 
background 
\ep flux within the beampipe, allowing the vertex detector to be placed
close to the beamline, with the first layer perhaps as close as 1cm.

\subsection{Tracking System} 
\vspace*{-.2cm}
A large-volume time projection chamber (TPC) offers high effective 
spatial granularity and yields low occupancy in the expected 
background $\gamma$ flux.
With a compensating coil to achieve $\Delta B/B$ $\simeq$ 0.2\% a
momentum resolution $\Delta p_t/p_t$ $\simeq$ $4.5\times10^{-5}$ (4T)
could be achieved. A wire-chamber TPC readout offers a useful degree
of particle identification, with $\pi/K$ separation up to or beyond $p$ = 20 GeV/$c$. 
Other possible readout technologies, such as gas electron multipliers and
micromesh gaseous structures, are under active development.
An `intermediate' tracker would provide linking hits between the VXD and TPC; 
two planes of $50\times500\mu$m$^2$ pixels appear to be sufficient.
Several planes of silicon pixels ($50\times200\mu$m$^2$) and crossed strips 
($50\times25\mu$m$^2$) are also planned to improve
tracking performance in the forward regions $7^{\circ}<\theta < 30^{\circ}$
from the beamline.

\subsection{Calorimetry} 
\vspace*{-.2cm}
Currently thinking is to put the electromagnetic and main
hadron calorimeters inside the solenoid. These would be
roughly 25 $X_0$ and 5-6 $\lambda_0$ thick, respectively.
Several options are being considered: a `tile' calorimeter based
on a sandwich of absorber and scintillator with wavelength-shifting fibres
taking the signal out; the same materials, but in a `Shashlik' (nearly
longitudinal fibres) configuration; and a `high-granularity' calorimeter
based on highly-segmented W/Si or Pb/scintillator or Pb/Ar for the
electromagnetic part and W, Fe or Pb with gas chambers for the hadronic
part. The first two options would have $\sim$ $3\times3$cm$^2$ transverse
segmentation and coarse longitudinal segmentation, with O($10^5$) channels; the
third option might have $1\times1$cm$^2$ transverse segmentation with 
readout of every layer, yielding as many as $10^7$ channels. All three options
are roughly comparable in terms of their energy resolution: 10\%/$\sqrt{E}$
(or better) (EM) and 40\%/$\sqrt{E}$ (had). The high-granularity option offers
additionally superb photon and neutral hadron identification capability.

\subsection{Muon System} 
\vspace*{-.2cm}
Space limitations prevent an adequate description, but an instrumented
return yoke is being considered for the muon identification and tracking
system, as well as to provide a tail-catcher hadron calorimeter.
Iron instrumented with gas chambers, such as resistive plate chambers
or streamer tubes, provides a well-tested and robust technology base for
strip and pad readout systems.

\section{Future Milestones}
\vspace*{-.2cm}
A detailed technical design report for both the
TESLA collider and detector is in preparation, and will be presented
in spring 2001, with subsequent evaluation by the German Science
Council. A decision on this multinational
project might be made as early as 2002, with construction
starting in 2003. This would allow turn-on in 2009, just a
few years after the startup of the LHC, providing a powerful partnership
for exploration of new physics.

\begin{figure}[h]
\vspace*{-0.3cm}
\begin{center}
\hspace*{.7cm}
\epsfxsize=12cm
\epsfysize=10cm
\epsffile{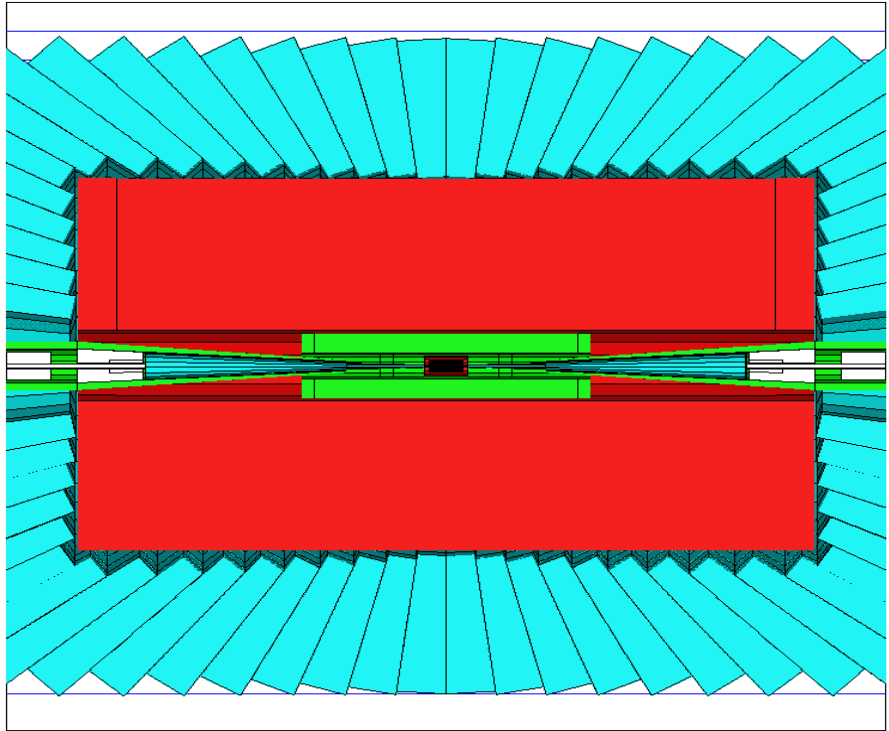}
\vspace*{-0.7cm}
\caption{
Current detector layout; scale: $6\times6$m$^2$.
}
\end{center}
\label{fig:det}
\vspace*{-0.8cm}
\end{figure}

\end{document}